\documentclass[aps,prc,showpacs,twocolumn,superscriptaddress]{revtex4}
\usepackage{graphicx}
\usepackage[dvips]{color}
\usepackage{colordvi}
\usepackage{bm}
\usepackage{amsmath}
\begin{document}
\bibliographystyle{apsrev}

\title{Thermodynamic properties of nuclear matter with three-body 
forces}
%\thanks{Research supported in 
% part by the Polish Ministry of Science and Higher Education, 
% grant N N202 1022 33}
\author{     V. Som\`a
}
\thanks{Electronic address~:
vittorio.soma@ifj.edu.pl}
\affiliation{
Institute of Nuclear Physics PAN, PL-31-342 Krak\'ow, Poland}

\author{     P. Bo\.{z}ek}
\thanks{Electronic address~:
piotr.bozek@ifj.edu.pl}
\affiliation{
Institute of Physics, Rzesz\'ow University, PL-35-959 Rzesz\'ow, Poland}
\affiliation{
Institute of Nuclear Physics PAN, PL-31-342 Krak\'ow, Poland}

\date{\today}

\begin{abstract}
We calculate thermodynamic quantities in symmetric
nuclear matter within the
self-consistent Green's functions method including three-body forces.
The thermodynamic potential is computed directly from a diagrammatic
expansion, implemented with the CD-Bonn and Nijmegen nucleon-nucleon
potentials and the Urbana three-body forces.
We present results for entropy and pressure up to temperatures of $20 \,
\mbox{MeV}$ and densities of $0.32 \, \mbox{fm}^{-3}$.
While the pressure is sensitive to the inclusion of three-body forces, 
the entropy  is not.
The unstable spinodal region is identified and the critical temperature
associated to the liquid-gas phase transition is determined.
When three-body forces are added we find a strong reduction of the critical
temperature, obtaining $T_c \simeq 12 \, \mbox{MeV}$.
\end{abstract}


\maketitle

The determination of thermodynamic properties of hot nuclear matter is
extremely relevant for its applications in astrophysics and heavy-ion reactions.
Pressure and entropy at finite temperature are crucial ingredients in the
modeling of core-collapse supernovae and protoneutron stars, which involve
processes characterized by densities up to several time the nuclear saturation
density and temperatures up to few tens of MeV's.
The nuclear equation of state (EOS) plays an important role also in the interpretation of
nucleus-nucleus collisions, in which a hot and dense state of matter is formed.
Of particular interest is the study of a possible phase transition occurring
at subsaturation densities. Because of the van der Waals nature of
nucleon-nucleon (NN) interactions, at low densities and high temperatures
nuclear matter is expected to undergo a first-order transition to a gas phase
\cite{Bertsch:1983uv}. 
It is claimed that an evidence for such behavior is found in intermediate
energy heavy ion collisions, as a plateau of the caloric curve derived for
the light fragments as a function of the reaction energy \cite{Natowitz:2002}.

The liquid-gas transition in nuclear matter presents some differences
with respect to the case of finite nuclei where, as the temperature increases,
Coulomb forces together with the decrease of the surface tension trigger the
onset of mechanical instabilities. 
One can try to relate 
the limiting temperature of nuclei $T_l$ to the critical temperature
in infinite matter $T_c$ by properly taking into account Coulomb and surface
effects \cite{Levit:1985zz}.
Different estimates from effective models based
on Skyrme forces \cite{Natowitz:2002prl} and more microscopic approaches 
\cite{TerHaar:1986fg, Baldo:2004wc}
are in agreement and yield a ratio $T_l/T_c$ of about $1/3$.

A reliable many-body theory is called for the interpretation of the heavy ion
reactions in the goal of extracting the equation of state and the critical
point of the nuclear liquid-gas phase transition \cite{Chomaz:2003dz} 
and in order to extrapolate to higher
densities and arbitrary isospin asymmetries. 
There exist only few realistic calculations of the nuclear matter EOS at
finite temperature. The variational approach, which yields reliable results at
$T=0$, is usually extended to finite temperatures by neglecting the
modifications of the correlation functions \cite{Friedman:1981qw}.
There are attempts of developing a finite $T$ variational technique 
\cite{Mukherjee:2006rt}
but no calculations of thermodynamic quantities are available so far.
The Br\"uckner-Hartree-Fock method generalized by means of the Bloch-de
Dominicis formalism has been applied to finite temperature nuclear matter\cite{Baldo:1998ah, Zuo:2006wx}.
The Green's functions in-medium T-matrix \cite{Dickhoff:1998zz,Bozek:1998su}
 approach is suitable for computing consistently
microscopic properties and thermodynamic observables and naturally takes into
account finite temperature correlations 
\cite{KadanoffBaym}. 
Thermodynamic relations 
such as the Hugenhotlz-van Hove and Luttinger identities
\cite{Hugenhotlz:1959, Luttinger:1960}
are automatically fulfilled by the $\Phi$-derivable T-matrix approximation
\cite{Baym:1962sx,Bozek:2001tz}.

In two recent papers we presented the first calculations of 
entropy and pressure
at finite temperature within the thermodynamically consistent T-matrix
approach \cite{Soma:2006}
and the first results which include three-body forces in the finite
temperature Green's functions method in nuclear matter \cite{Soma:2008nn}.
Following these works we compute here pressure and entropy without and with
three-body forces in the case of symmetric nuclear matter for two
different realistic NN potentials, in particular addressing the unstable
region related to the liquid-gas phase transition.
\\
\\

Let us briefly recall the adopted computational method, whose details can be
found in Refs. \cite{Soma:2006, Soma:2008nn}. We solve the set of coupled
equations which involve the calculation of the T-matrix, summing 
the ladder diagrams at all orders, the determination of the
nucleon self-energy and the single-particle Green's function.
The full off-shell
propagation of particles in the medium is taken into account.
We employ two different realistic nucleon-nucleon interactions, the CD-Bonn
\cite{Machleidt:2000ge} and the Nijmegen \cite{Stoks:1994wp} parameterizations, 
together with the semi-microscopic Urbana
three-nucleon potential \cite{Carlson:1983}.
Three-body forces, necessary for a reliable description of the saturation
properties in symmetric nuclear matter, are included via an effective two-body
interaction derived after averaging out the third particle. The effect on the
two remaining nucleons results in a mean field whose two parameters
are fixed by requiring the correct saturation density and binding energy.
The range of temperatures and densities studied is well above the superfluid 
transition in symmetric nuclear matter \cite{Bozek:2002jw}.

\begin{figure}[h!]
\begin{center}
\includegraphics[width=9cm]{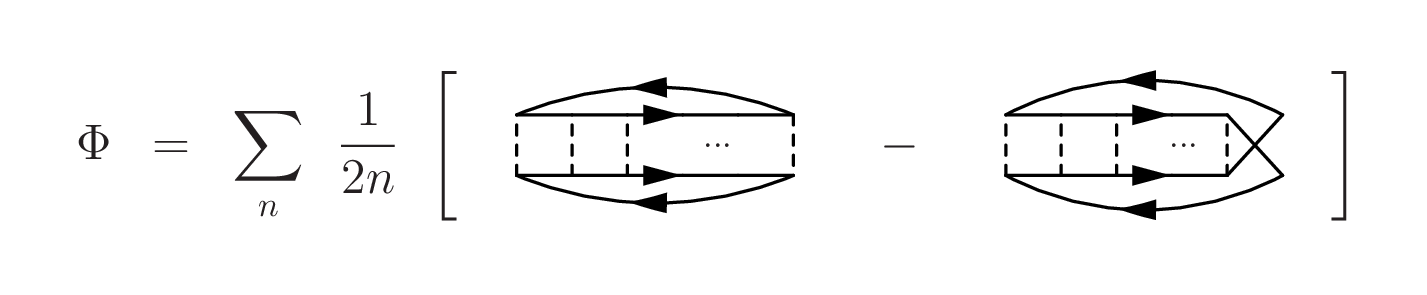}
\caption{Diagrammatic expansion of the generating functional $\Phi$.
The sum includes ladder-type diagrams up to infinite order
($n$ is the number of interaction lines in each diagram).}
\label{fig:phi}
\end{center}
\end{figure}

After calculating the single-particle propagator $G$ and the nucleon
self-energy $\Sigma$ from the iterative scheme
of the T-matrix equations one can compute the grand canonical potential,
expressed as
\begin{equation} 
\label{om}
\Omega=-\mbox{Tr}\{\ln[G^{-1}]\} -\mbox{Tr}\{\Sigma G\} + \Phi \: .
\end{equation}
The generating functional $\Phi$ has a diagrammatic expansion similar
to the one  for the interaction energy in the T-matrix approximation, differing
by a factor $1/n$ ($n$ is the number of interaction lines) in front of each
diagram (see Fig. \ref{fig:phi}).
Therefore we derive $\Phi$ from the expression for 
$\langle H_{int} \rangle$ by multiplying the
potential $V$ by a numerical factor $\lambda$ and integrating
\begin{equation}
\label{eq:ourphi}
\Phi=\int_0^1
\frac{d\lambda}{\lambda} \, \langle H_{int}(\lambda V, G_{\lambda=1}) 
\rangle \: ,
\end{equation}
where $G$ is the dressed single-particle propagator computed in the system
with the full strength interaction.
The expectation value $\langle H_{int} \rangle$ itself can be easily obtained
from the $T$ matrix \cite{Soma:2006}.
From the grand canonical potential (dividing by the volume
$\mathcal{V}$) we obtain the pressure
\begin{equation} 
\label{eq:p2}
P = - \frac{\Omega}{\mathcal{V}} \: ,
\end{equation}
which is then computed directly from its diagrammatic expansion
without the use of numerical derivatives.

\begin{figure}[h!]
\begin{center}
\includegraphics[width=8.5cm]{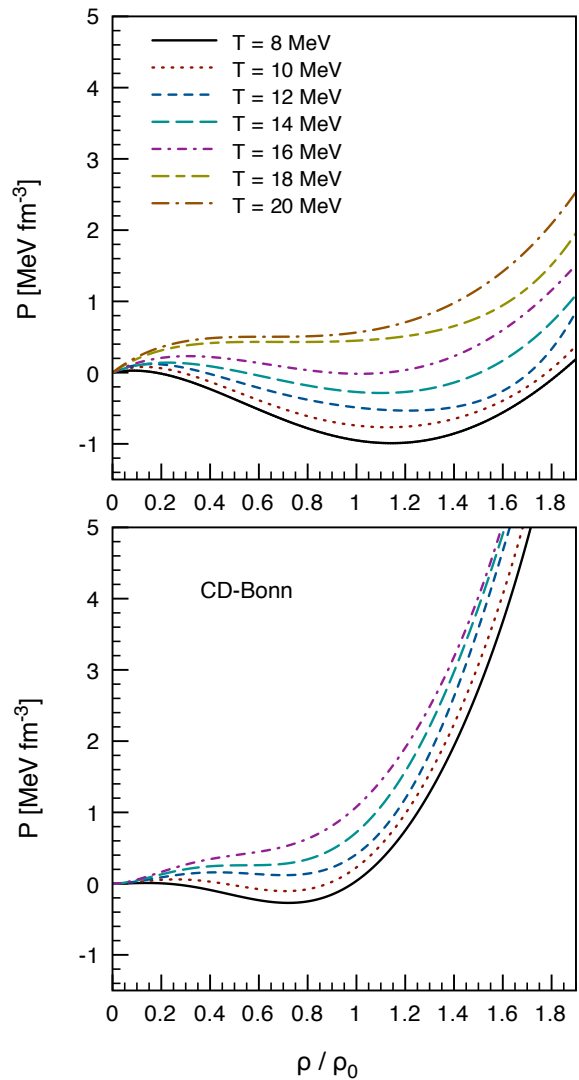}
\caption{(Color online) 
Pressure as  function of density in symmetric nuclear matter
for different temperatures
with the CD-Bonn potential (upper panel) 
and the CD-Bonn potential
plus three-body forces (lower panel).}
\label{fig:pr-cdb}
\end{center}
\end{figure}

\begin{figure}[h!]
\begin{center}
\includegraphics[width=8.5cm]{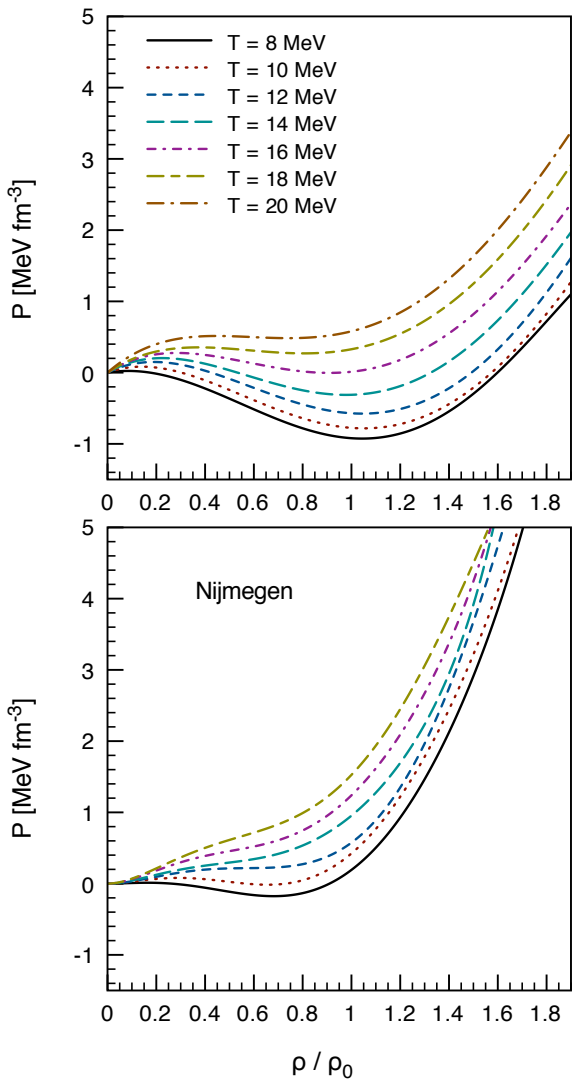}
\caption{(Color online) 
Pressure as  function of density in symmetric nuclear matter
for different temperatures
with the Nijmegen potential (upper panel) and the Nijmegen potential
plus three-body forces (lower panel).}
\label{fig:pr-nij}
\end{center}
\end{figure}

In Figs. \ref{fig:pr-cdb} and \ref{fig:pr-nij}
we present the pressure  for symmetric nuclear matter 
without and with
three-body forces for the CD-Bonn and for the Nijmegen potential
respectively. 
Results are shown for different temperatures as  function of the density.
When looking at the curves obtained with the two-body NN potentials
(upper panels) we notice that they cross the thermodynamically forbidden
region in which the derivative of the pressure with respect to
the density is negative.
For both CD-Bonn and Nijmegen interactions the inclusion of
three-body forces (lower panels)
leads to a stiffening of the dependence of the pressure on density. 
The spinodal instability region below saturation density 
is still present but gets substantially reduced.

In order to address more in details the critical behavior and 
the effect of three-body forces on the critical temperature for the
liquid-gas phase transition we study the limits of the instability region
in the density-temperature plane.
The spinodal region is characterized by the negative
derivatives of the pressure and the chemical potential
\begin{equation}
\label{eq:liq-gas-r}
\left. \frac{\partial \, P}{\partial \, \rho} \right|_T < 0 \: ,\hspace{1.6cm}
\left. \frac{\partial \, \mu}{\partial \, \rho} \right|_T < 0 \: .
\end{equation}
Inside this region the system is unstable and tends to separate in two 
different phases, gas at a lower and liquid at a higher density. 
The two phases may coexist in an interval of densities and temperatures
up to the point in which
\begin{equation}
\label{eq:liq-gas-coex-p}
P(\rho_{\, \footnotesize{\mbox{gas}}}) = 
P(\rho_{\, \footnotesize{\mbox{liquid}}}) 
\end{equation}
and
\begin{equation}
\label{eq:liq-gas-coex-mu}
\mu(\rho_{\, \footnotesize{\mbox{gas}}}) = 
\mu(\rho_{\, \footnotesize{\mbox{liquid}}})  \: .
\end{equation}
The two regions end up in a coincident point which defines the critical
density and temperature of the liquid-gas phase transition.
\begin{figure}[h!]
\begin{center}
\includegraphics[width=8cm]{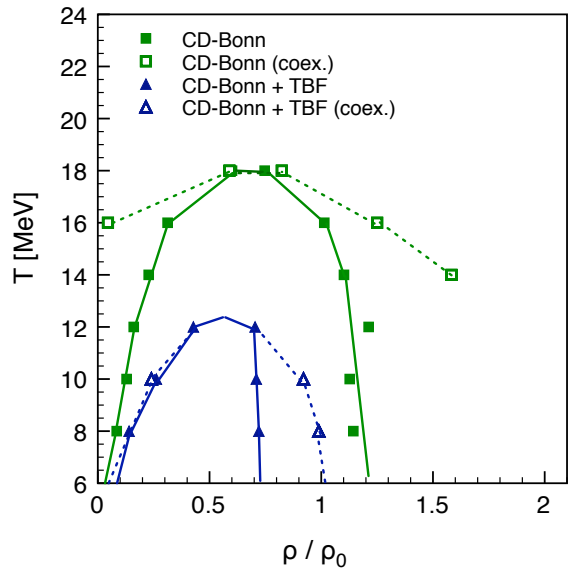}
\caption{(Color online) 
Coexistence and spinodal lines for the CD-Bonn potential
without and with three-body forces.}
\label{fig:lg-cdb}
\end{center}
\end{figure}
\begin{figure}[h!]
\begin{center}
\includegraphics[width=8cm]{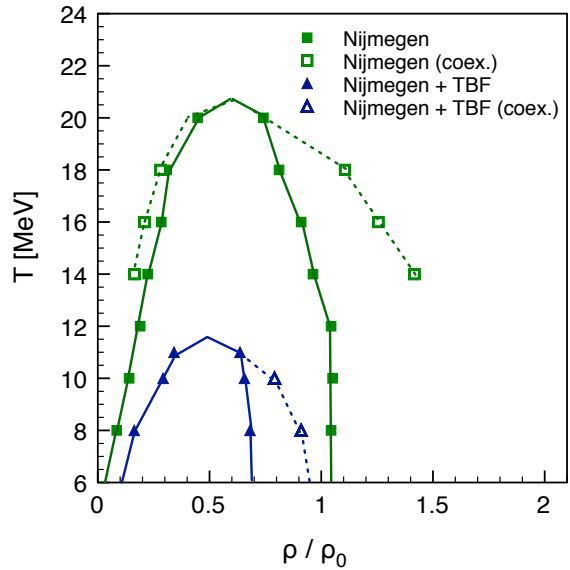}
\caption{(Color online) 
Coexistence and spinodal lines for the Nijmegen potential
without and with three-body forces.}
\label{fig:lg-nij}
\end{center}
\end{figure}

We compute the spinodal and the coexistence lines for the CD-Bonn and the
Nijmegen potential without and with the inclusion of three-body forces.
The limits of the spinodal region defined by Eq. 
\eqref{eq:liq-gas-r} are found as
local maxima and minima of the pressure.
The conditions on the pressure and chemical potential
are equivalent for a one-component system
such as symmetric or pure neutron matter. For the case of arbitrary
isospin asymmetries one should in principle consider both of them,
but there is evidence that chemical and mechanical
instabilities coincide also for asymmetric nuclear matter
\cite{Margueron:2002wk}.

We use two methods for determining the spinodal line: 
we compute the pressure derivative 
directly from the interpolation of $P$ and apply the one 
suggested by Baldo and Ferreira \cite{Baldo:1998ah}
which consists in plotting the chemical potential as a function of the pressure,
looking for the back bending of the curves.
The results obtained with the two techniques do coincide,
confirming the stability of the calculations even in the low density
regime.
The critical lines are displayed in Figs. \ref{fig:lg-cdb} and \ref{fig:lg-nij}
for CD-Bonn and Nijmegen respectively. At low temperatures the unstable
and coexistence phases are present over a large range of densities. As the 
temperature increases the instability region becomes smaller ending up with
the point defining the critical temperature for the liquid-gas transition.
The critical temperature is
$T_c = 18 \, \mbox{MeV}$ for the CD-Bonn and 
$T_c = 20.5 \, \mbox{MeV}$ for the Nijmegen
potential. Both the spinodal and
the coexistence regions
get strongly reduced when three-body forces are included, with the
critical point at $T_c = 12.5 \, \mbox{MeV}$ when CD-Bonn 
is employed and 
$T_c = 11.5 \, \mbox{MeV}$ if we consider the Nijmegen
potential.

Another quantity which characterizes the liquid-gas phase transition is 
a dimensionless parameter that comprises the critical pressure,
density and temperature,
$P_c/(\rho_c \, T_c)$, which assumes the value
3/8 for a van der Waals equation of state. 
\begin{table}[h!] 
\begin{center} 
\begin{tabular}{|c||c|c|c|c|} 
\hline  potential & $T_c \, (\mbox{MeV})$ & 
$ \rho_c \, (\mbox{fm}^{-3})$ &
$ P_c \, (\mbox{MeV} \, \mbox{fm}^{-3}) $& 
$ \displaystyle \, \frac{P_c}{\rho_c \, T_c} \, $
\\ 
 \hline 
\hline  CD-Bonn  & 
$18$ & $0.107$ & $0.43$ & 
$0.22$  
\\ 
\hline  CD-Bonn + TBF & 
$12.5$ & $0.096$ &  $0.14$ & 
$0.12$ 
\\ 
\hline  Nijmegen & 
$20.5$ & $0.094$ & $0.50$ & 
$0.26$  
\\
\hline Nijmegen + TBF & 
$11.5$ & $0.088$ & $0.15$ & 
$0.14$  
\\
\hline
\end{tabular} 
\end{center} 
\caption{Critical values of the  temperature, density and pressure of the
liquid-gas phase transition for the CD-Bonn and Nijmegen potentials
without and with three-body forces.
}
\label{tab:prt} 
\end{table}
In Table \ref{tab:prt} we summarize these critical quantities for the
various potentials without and with three-body forces. The values
for the CD-Bonn and the Nijmegen potentials are rather similar, with
a strong decrease of the dimensionless parameter after the inclusion of
three-body forces, signaling a departure from the van der Waals
equation of state.

Our estimates with the two-body nucleon-nucleon interactions only 
are in agreement with other microscopic calculations.
Rios et al. \cite{Rios:2008ef} use a self consistent Green's functions
approach but compute the free energy from the Carneiro-Pethick
quasiparticle entropy \cite{Carneiro:1975},
obtaining for the CD-Bonn potential a critical temperature
$T_c = 18.5 \, \mbox{MeV}$ and $P_c/(\rho_c \, T_c) = 0.20$.
When using the Argonne V18
parameterization, however, they find that $T_c$  is reduced to about
$11 \, \mbox{MeV}$ and $P_c/(\rho_c \, T_c) = 0.14$. 
Baldo and Ferreira 
\cite{Baldo:1998ah} performed calculations within
the Bloch-de Dominicis finite temperature generalization of
the Br\"uckner-Hartree-Fock method. Using the 
Argonne V18 potential they estimate the critical temperature to be
$T_c \simeq 21 \, \mbox{MeV}$.
However, when they include the Urbana three-nucleon potential in the
calculations they find, in contrast with our result, that three-body forces
do not strongly affect the critical temperature, which is reduced
to $T_c \simeq 20 \, \mbox{MeV}$.
Other calculations within the Bloch-de Dominicis formalism
\cite{Das:1992hn} or the relativistic Dirac-Br\"uckner-Hartree-Fock
approach \cite{TerHaar:1986fg, Huber:1998dd}, on the other hand, yield
lower values of $T_c$, respectively 
$9, 12 \, \mbox{and} \, 10.4 \, \mbox{MeV}$, 
closer to our result with three-body forces.

\begin{figure}[h!]
\begin{center}
\includegraphics[width=8cm]{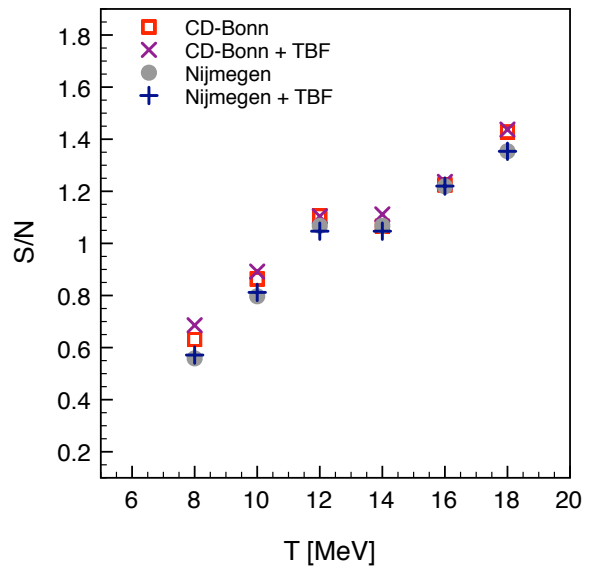}
\caption{(Color online) Entropy per particle for symmetric nuclear matter
at $\rho = \rho_0$ 
with the CD-Bonn and Nijmegen potentials without and with
three-body forces.}
\label{fig:ent}
\end{center}
\end{figure}

We compute the entropy per particle in the interacting system
from the thermodynamic relation
\begin{equation}
\label{eq:ent}
\frac{S}{N} = \frac{1}{T} 
\left [
\frac{E}{N} + \frac{P}{\rho} - \mu
\right ] \: .
\end{equation}
The results for symmetric nuclear matter at $\rho = \rho_0$ are shown
in Fig. \ref{fig:ent} for different temperatures.
The effects of three-body forces are very small, as well as the dependence
on the nucleon-nucleon potential. The entropy appears to be independent
of the details of the interaction, supporting the conclusion that it is not
much affected by nucleon correlations. 
The reliability of calculations of the entropy 
at the level of a quasiparticle approximation has been checked also by
Rios et al. \cite{Rios:2006nr} using the Carneiro-Pethick dynamical
quasiparticle formula \cite{Carneiro:1975}.
\\
\\

Existing estimates of the critical temperature for the
liquid-gas transition in infinite nuclear matter are in the
range $T_c \simeq 9 - 21 \, \mbox{MeV}$. Within the finite temperature
Green's functions approach we calculate thermodynamic properties
of symmetric matter and study the spinodal instability region related to
the first-order phase transition. We observe that three-body forces have a strong
effect on the pressure of the interacting system, reducing the size of the
unstable region in the density-temperature plane. 
With the CD-Bonn and the Nijmegen potential we find that the critical
temperature decreases respectively from $T_c = 18 \, \mbox{MeV}$
to $T_c = 12.5 \, \mbox{MeV}$ and from $T_c = 20.5 \, \mbox{MeV}$
to $T_c = 11.5 \, \mbox{MeV}$. In all cases the critical density lies in the
range $\rho_c \simeq 0.09 - 0.11\mbox{fm}^{-3}$. The pressure
at which the phase transition takes place decreases from
$P_c \simeq 0.43 - 0.50 \, \mbox{MeV}\,\mbox{fm}^{-3}$ to
about $P_c = 0.15 \, \mbox{MeV}\,\mbox{fm}^{-3}$ after the inclusion
of the three-nucleon potential.

Three-body forces do not affect the entropy, which turns out to be independent
of the potential used in the calculations. This last results 
 confirms that
entropy is not much sensitive to nucleon-nucleon correlations
(cf. Refs. \cite{Soma:2006, Rios:2006nr}).

\section*{Acknowledgements}

Research supported in
  part by the Polish Ministry of Science and Higher Education, 
  grant N N202 1022 33.
V. S. acknowledges the Academic Computer Center CYFRONET of the
AGH University of Science and Technology in Krak\'ow for benefiting
from the use of its high performance computers.

%\clearpage

\bibliography{vibi}

\end{document}